\documentclass{elsart}
\pdfoutput=1
\usepackage[dvips]{graphics}
\usepackage{graphicx}
\usepackage{amsfonts}
\usepackage{amssymb}
\usepackage{amsmath}
\usepackage{subfigure}
\usepackage{color}

\begin{document}

\begin{frontmatter}

\title{Dark-Bright Discrete Solitons: A Numerical Study of Existence,
Stability and Dynamics}

\author[US]{A. \'Alvarez},
\author[EUP]{J. Cuevas\thanksref{cor}},
\author[US]{F.R. Romero},
\author[UMA]{P.G. Kevrekidis}

\address[US]{Grupo de F{\'i}sica No Lineal, Universidad de
Sevilla. \'Area de F\'{\i}sica Te\'orica. \\
Facultad de F\'{\i}sica, Avda. Reina Mercedes, s/n, E-41012
Sevilla, Spain}
\address[EUP]{Grupo de F{\'i}sica No Lineal, Universidad de
Sevilla. Departamento de F{\'i}sica Aplicada I. \\ Escuela
Polit{\'{e}}cnica Superior, C/ Virgen de {\'{A}}frica, 7, E-41011
Sevilla, Spain}
\address[UMA]{Department of Mathematics and Statistics, University of
Massachusetts, Amherst MA 01003-4515, USA}
\thanks[cor]{Corresponding author.
             E-mail: jcuevas@us.es}

\begin{abstract}
In the present work, we numerically explore the existence and
stability properties of different types of configurations of
dark-bright solitons, dark-bright soliton pairs and pairs of
dark-bright and dark solitons in discrete settings, starting from
the anti-continuum limit. We find that while single discrete dark-bright
solitons have similar stability properties to discrete dark solitons,
their pairs may {\it only} be stable if the bright components are
in phase and are always unstable if the bright components are out of
phase. Pairs of dark-bright solitons with dark ones have similar
stability properties as individual dark or dark-bright ones.
Lastly, we consider collisions between dark-bright solitons and
between a dark-bright one and a dark one. Especially in the latter
and in the regime where the underlying lattice structure matters,
we find a wide range of potential dynamical outcomes depending on
the initial soliton speed.
\end{abstract}

\journal{Physica D}

\date{\today}

\begin{keyword}
Solitons \sep
Dark solitons \sep
Discrete Nonlinear Schr\"odinger equations \sep
Manakov model.

\end{keyword}

\end{frontmatter}

\section{Introduction}

The setting of multi-component dispersive systems of nonlinear
waves is one which has had a significant impact on a variety
of areas including the nonlinear optics of fibers and crystals~\cite{yuri}
and the recently blossiming area of Bose-Einstein condensates (BECs)
in atomic physics~\cite{stringarius}. A fascinating example
among the complex nonlinear localized structures that can
emerge in such settings consists of the so-called
symbiotic solitons. These are waveforms that would {\it not}
otherwise exist in unary systems, but are nevertheless
supported in multi-component ones, precisely because of the
inter-species interaction (be they atomic species in BEC or
different frequencies, polarizations or other optical ``species''
in nonlinear optics). One of the principal examples of this type
consists of the dark-bright (DB) solitons in two-component systems
with the self-defocusing nonlinearity. The latter class of
nonlinearities is well-known to support dark solitary wave
structures~\cite{luther}. However, in the presence of nonlinear
interspecies interactions, these dark solitons form a localized
potential which allows the trapping in the form of bound states
of bright-soliton-like, density bumps leading to the formation
of DB waves.

 Remarkably, such DB structures have been experimentally
monitored both in nonlinear optics and in atomic physics.
They were first experimentally created in
photorefractive crystals through the pioneering experiments of \cite{seg1} and
subsequently their interactions were partially monitored in
\cite{seg2}\footnote{It should be noted that theoretical
proposals of symbiotic DB solitons (although in a somewhat different
setting than what will be considered below) have existed much earlier
and
at least since the work of \cite{ysk91}.}.
More recently, the realization of multi-component atomic
BECs~\cite{Myatt1997a,Hall1998a,Stamper-Kurn1998b} has prompted a
number of theoretical investigations of DB waves in the latter
setting as well starting with the work of~\cite{buschanglin},
which examined the trapped dynamics of DB solitons in the presence
of the parabolic traps relevant to BECs. Subsequently,
these structures were also extended to more complex
settings such as the spinor three-component system, where
dark-dark-bright or bright-bright-dark solitons could
arise~\cite{DDB}. More recently, the interaction between such dark-bright
solitons~\cite{rajendran,berloff} and their potential higher-dimensional
generalizations into symbiotic, so-called, vortex-bright solitons~\cite{VB}
have been also considered. Perhaps more importantly, the experiments
of~\cite{hamburg} were able to provide the first observations
of such DB solitons and their interactions e.g. with another
dark soliton. A realization of multiple dark-bright solitons
has also emerged in the experiments of \cite{enge1} and the
oscillations of dark-bright solitons in a trap, as well as the
interaction between two dark-bright solitons were experimentally
monitored in \cite{enge2}.

On the other hand, an area of studies that has also received
significant attention both within the realm of nonlinear
optics and within that of atomic physics concerns the study
of nonlinear dynamical lattices. In these, the ``evolution variable''
is continuum, while the ``spatial variables'' are discrete.
One of the principal reasons for the growth of this field has
been the development of optically induced lattices in photorefractive media,
such as Strontium Barium Niobate
(SBN) \cite{efrem}, and their experimental realization \cite{moti1,moti2}. As a
result, a remarkable set of
nonlinear waves has been predicted and experimentally observed; these
have now been summarized in a number of reviews \cite{lederer,gorbach}.
In addition, another direction of nonlinear optics that has
grown in parallel (and has partially preceded the one above)
concerns the intriguing interplay of
nonlinearity and discrete diffraction emerging in fabricated AlGaAs
waveguide arrays \cite{7}. The numerous interesting phenomena
observed therein including discrete diffraction, Peierls barriers,
diffraction management, gap solitons, vector solitons, and modulational
instabilities among others have now been summarized e.g. in
\cite{review_opt,lederer,gorbach}. Lastly, over the last
decade, nonlinear dynamical lattices have emerged as a theme of
interest in BECs since droplets of such condensates
can be trapped in periodic optical potentials often referred to as
``optical lattices'' (see, e.g., \cite{ibloch} and references therein).
Among the principal byproducts of these studies have been
the manifestation of dynamical instabilities, Bloch oscillations,
Landau-Zener tunneling, and gap solitons as has been summarized
in the reviews~\cite{konotop,markus2}.

Although, to the best of our knowledge dark-bright solitons
have {\it not} been realized in dynamical lattice settings,
all of the relevant ingredients illustrating their experimental
tractability are available. Namely, recent nonlinear optical
experiments have
considered the realization of dark solitons in the presence
of self-defocusing nonlinearities of both
the Kerr type in AlGaAs waveguide arrays~\cite{silb} and
of the saturable type due to the photovoltaic effect in
lithium niobate arrays~\cite{kip}. On the other hand,
vector dynamical lattice systems have also been recently
experimentally monitored; see e.g.~\cite{christo} and the discussion
therein.

From the above background, it becomes clear that the investigation
of DB waveforms in discrete two-component self-defocusing systems is a
timely and relevant theme. Our scope in the present work is
to numerically explore this topic and the corresponding existence,
stability and collision dynamics of DB states in nonlinear dynamical
lattices starting from a well-established limit, namely the
anti-continuum one of~\cite{mackay}. The latter has been extremely
helpful in deriving properties of bright and dark discrete solitons
in both focusing and defocusing lattices, as can be seen in
the detailed discusssion of~\cite{pgk}. In what follows, we will
see that some of the DB states (such as a single DB soliton
or pairs of DB solitons with dark ones) draw significant parallels
with previously studied cases such as single-component discrete
dark solitons. On the other hand, we will also show that other
states such as standing wave pairs of two dark-bright solitons
have some surprising properties which defy the single-component
intuition. As will be seen the linear stability of such
states will be critically affected by the relative phase of
the bright components. Lastly, we will examine the collisional
dynamics of two DB solitons, as well as that of a DB state
with a dark one. The latter will reveal a significant wealth of
potential outcomes depending on a single parameter, namely the
wave speed. Our presentation will be structured as follows.
In section 2, we present the relevant prototypical model setup
in the form of two coupled defocusing discrete nonlinear
Schr{\"o}dinger (DNLS) equations.
Then, in section 3, we explore the existence and stability
properties of the DB solitons. In section 4, we
consider their collisional dynamics. Finally, in section 5,
we summarize our findings and present our conclusions.

\section{Model Setup}\label{sec:model}
We consider the system of two coupled discrete nonlinear
Schr{\"o}dinger (CDNLS) equations given by~\cite{pgk}:
\begin{eqnarray}\label{dyn1}
    i\dot U_n+(g_{11}|U_n|^2+g_{12}|V_n|^2)U_n+C\Delta U_n &=& 0
\nonumber\\
    i\dot V_n+(g_{12}|U_n|^2+g_{22}|V_n|^2)V_n+C\Delta V_n &=& 0,
\end{eqnarray}

where $\Delta$ is the discrete Laplacian operator($\Delta U_n=U_{n+1}+U_{n-1}-2U_n$).

Our focus in what follows will be on stationary solutions of the system of
equations~(\ref{dyn1}). To that effect we use the standard \textsl{ansatz}

\begin{equation}\label{ans2}
    U_n(t)=\exp(-i\Lambda_1t)u_n \qquad ;\qquad
    V_n(t)=\exp(-i\Lambda_2t)v_n,\\
\end{equation}

with frequencies (propagation constants in nonlinear optics or chemical
potentials in BECs) $\Lambda_1$ and $\Lambda_2$, and spatial profiles
$\{u_n\}$ and $\{v_n\}$.
Using (\ref{ans2}) in (\ref{dyn1}), we obtain the system of coupled difference
 equations
\begin{eqnarray}\label{dyn2}
    \Lambda_1 u_n+(g_{11}|u_n|^2+g_{12}|v_n|^2)u_n+C\Delta u_n &=& 0
    \nonumber\\
    \Lambda_2 v_n+(g_{12}|v_n|^2+g_{22}|v_n|^2)v_n+C\Delta v_n &=& 0.
 \end{eqnarray}
The anti-continuum (AC) limit of~\cite{mackay} (see also~\cite{pgk})
concerns the case with $C=0$. In the absence of the potential, the
system of Eq. (\ref{dyn2}) results into a pair of analytically solvable
algebraic equations for each site.
In what follows, and given the nature of the DB
solitons, we will restrict our attention to cases where the first
(dark) component is excited and the second (bright) one is not
i.e., $u_n=\pm \sqrt{-\Lambda_1/g_{11}}$, $v_n=0$ (at the tails of the
DB soliton) or to ones where the first component is inert and
the second one is excited i.e., $u_n=0$ and $v_n=\pm \sqrt{-\Lambda_2/g_{22}}$.
(at the center of the DB soliton). Once an exact solution of the DB type
has been constructed at the AC limit, it can be continued (numerically,
as well as analytically) for finite values of the coupling $C$.
It is worth remarking that throughout the text, unless otherwise
specified, free ends boundary conditions have been used.

A relevant question concerning the solutions at finite values of $C$
is that of dynamical stability. To that effect, we use linear stability
analysis to provide a spectral response to this question. The stability is determined in a frame rotating with frequency
$\Lambda_1$ for $U_n(t)$ and $\Lambda_2$ for $V_n(t)$, i.e., we
suppose that \cite{Interlaced}
\begin{equation}
    U_n(t)=\exp(-i\Lambda_1t)[u_n+\xi^{(1)}_n(t)],\qquad
    V_n(t)=\exp(-i\Lambda_2t)[v_n+\xi^{(2)}_n(t)].
\end{equation}

The small perturbations $\xi^{(k)}_n(t)$, with $k=1,2,$ can be
expressed as
\begin{equation}
    \xi^{(1)}_n(t)=a_n\exp(i\omega t)+b_n\exp(-i\omega^*t),\qquad
    \xi^{(2)}_n(t)=c_n\exp(i\omega t)+d_n\exp(-i\omega^*t),
\end{equation}
leading to the linear stability equations
\begin{equation}\label{evp}
\omega \left( \begin{array}{c} a_n \\ b_n^* \\ c_n \\ d_n^*
\end{array} \right)=\left(
\begin{array}{ccccc}
K_{1,n}& g_{11}u_n^2 & g_{12} u_nv_n^* & g_{12}u_nv_n\\
-g_{11}(u_n^2)^* & -K_{1,n} & -g_{12} u_n^*v_n^* & -g_{12}u_n^*v_n\\
g_{12} u_n^*v_n & g_{12}u_nv_n & K_{2,n} & g_{22}v_n^2\\
-g_{12} u_n^*v_n^* & -g_{12}u_nv_n^* & -g_{22}(v_n^2)^* & -K_{2,n}
\end{array}\right)\left( \begin{array}{c} a_n \\ b_n^* \\ c_n \\ d_n^*
\end{array} \right)
\end{equation}

with

\begin{eqnarray}
K_{1,n}&=&\Lambda_1+2g_{11}|u_n|^2+g_{12}|v_n|^2+\Delta,\nonumber\\
K_{2,n}&=&\Lambda_2+2g_{22}|v_n|^2+g_{12}|u_n|^2+\Delta.\nonumber
\end{eqnarray}

More specifically, it can be proved~\cite{pgk} that a solution is
spectrally
stable if all the $\omega$ eigenvalues (the so-called eigenfrequencies of the
system) are real numbers. Two type of instabilities are observed in
this system: (i) exponential instabilities, represented by a
eigenfrequency ($\omega$) pair with zero real part and non-vanishing
imaginary part and (ii) oscillatory instabilities, accounted for by a
Hamiltonian Hopf bifurcation, and represented by quartets of
eigenfrequencies with non-zero real and imaginary parts.

Throughout this paper we have assumed $g_{11}=-0.97$, $g_{22}=-1.03$ and $g_{12}=-1$, chiefly motivated by the experimental discussed interspecies
interaction ratios of~\cite{Myatt1997a,Hall1998a}. All the calculations have been performed assuming (without loss of generality) $\Lambda_1=1$, and for the frequency $\Lambda_2>0$ we have explored the range $\Lambda_2\in [0,1.5]$, as representative of the different parametric regimes.

\section{Stability properties of dark-bright multi-soliton
solutions}\label{sec:stability}

\subsection{Solutions at the anti-continuous limit}

In this subsection, we examine a representative set of structures which
are initialized at $C=0$. In the  AC limit, we use the following convenient notation to
represent our solutions: by indicating $\tilde{u}_n= \pm 1$, we
denote an excited site of $u_n= \pm \sqrt{-\Lambda_1/g_{11}}$, while
similarly for the second component $\tilde{v}_n=\pm 1$ is used to denote
$v_n= \pm \sqrt{-\Lambda_2/g_{22}}$. In this notation, we have
considered the following (AC-limit) solutions as seed points
for our numerical continuations:

\begin{itemize}

\item    Dark-bright one-soliton solutions (DB 1-S).

\begin{equation}
    \tilde{u}_n=\mathrm{sgn}(n),\qquad \tilde{v}_n=\delta_{n,0},
\end{equation}

where $\mathrm{sgn}(n)$ denotes the sign of $n$ and it is implied
that the solution assumes the value $\tilde{u}_0=0$.

\item  Dark-bright two-soliton solutions (DB 2-S). We have considered, as a representative case, that where the distance between the excited bright spots is of 2 sites:

\begin{equation}
    \tilde{u}_n=1-\delta_{|n|,1}-2\delta_{n,0}.
\end{equation}

In this setting, we distinguish two configurations for the bright part:

the in-phase bright case with

\begin{equation}
    \tilde{v}_n=\delta_{n,-1}+\delta_{n,1},
\end{equation}

and the out-of-phase (anti-phase) case with

\begin{equation}
    \tilde{v}_n=\delta_{n,-1}-\delta_{n,1}.
\end{equation}

\item  Dark-bright and dark two-soliton solutions (DB+D). Again, we have considered that there are two sites separating the vanishing sites of the $u_n$ field, one of which is associated with the dark-bright soliton, while the other is associated with the dark one.

\begin{equation}
    \tilde{u}_n=1-\delta_{|n|,1}-2\delta_{n,0},\qquad \tilde{v}_n=\delta_{n,-1}
\end{equation}

\end{itemize}

\begin{figure}
\begin{center}
\includegraphics[width=0.6\textwidth]{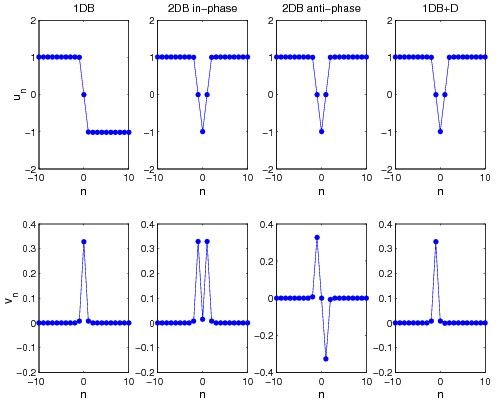}
\end{center}
\caption{Profile of the configurations analyzed in the paper. In all cases, $C=0.02$, $\Lambda_1=1$ and $\Lambda_2=0.15$.} \label{Fig0}
\end{figure}

Figure~\ref{Fig0} shows some examples of the profile of the above mentioned configurations.

It is interesting to note that at the AC limit, not only the
existence theory is transparent but also the nature of the
linearization spectrum. More specifically, notice that
all the DB states involve configurations for which $u_n v_n=0$,
hence the above $4 N \times 4 N$ (where $N$ is the
number of lattice sites considered) matrix splits into two diagonal
$2 N \times 2 N $ blocks for each of the components. The background
states (away from the DB center) provide each of these components
with a spectral dispersion relation of the form:
\begin{eqnarray}
\omega=\pm \sqrt{4 C \sin^2(\frac{q}{2}) \left(
4 C \sin^2(\frac{q}{2}) + 2 \Lambda_1 \right)}
\label{disp1}
\end{eqnarray}
between the linearization frequencies $\omega$ and wavenumbers $q$;
this relation stems from the dark (first) component. On the
other hand, the bright (second) component similarly yields
a dispersion relation for the background state (away from the center
of the DB)
\begin{eqnarray}
\omega=\pm \left(\frac{g_{12}}{g_{11}} \Lambda_1 - \Lambda_2 + 4 C
\sin^2(\frac{q}{2}) \right)
\label{disp2}
\end{eqnarray}
For the excited sites of the dark-bright soliton (where the
first component is vanishing i.e., dark and the second component
is non-vanishing i.e., bright), we obtain a pair of eigenfrequencies
satisfying:
\begin{eqnarray}
\omega=\pm \left(-\frac{g_{12}}{g_{22}} \Lambda_2 + \Lambda_1 \right)
\label{disp3}
\end{eqnarray}
from the dark component and a pair of zeros from the bright one.
The fundamental stability question concerns, of course, the fate
of these eigenfrequencies when $C \neq 0$ and is examined systematically
in our numerical computations within the following subsection.

\subsection{Numerical continuations for arbitrary coupling}

Having considered the existence and stability properties of the DB waveforms at the AC limit of $C=0$, we now turn to the case of
finite coupling $C$. In what follows, we examine for the
configurations
discussed in the previous subsection, the findings
pertaining to their existence when a continuation from $C=0$ is
performed, calculating
for each solution the eigenfrequencies stemming from the relevant
stability problem ~(\ref{evp}). In every case, $\Lambda_1=1$ is fixed
whereas $\Lambda_2$ is chosen in the interval $[0,1.5]$.

\subsubsection{Dark-bright one-soliton solutions}

\begin{figure}
\begin{center}
\includegraphics[width=0.6\textwidth]{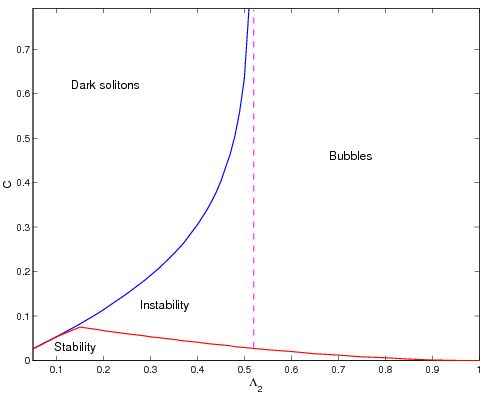}
\end{center}
\caption{Existence and  stability properties of the dark-bright one-soliton solutions as a function of coupling $C$ and frequency
$\Lambda_2$, for the case of $\Lambda_1=1$.} \label{Fig4}
\end{figure}

Three main behaviours are observed separated by the critical values $\Lambda_2=0.51$ and $\Lambda_2=1$. An overall view for the existence and stability properties of these solutions, as a function of coupling $C$ and frequency $\Lambda_2$, is illustrated in Fig.~\ref{Fig4}.

\begin{figure}
\begin{center}
\includegraphics[width=0.7\textwidth]{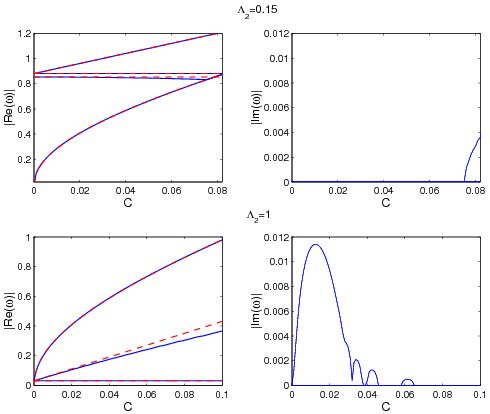}
\end{center}
\caption{Real part of the relevant spectral eigenfrequencies (left panels) and
  imaginary part of the eigenfrequency spectrum (right panels) versus $C$ of the
  DB one-soliton solutions. In both cases, $\Lambda_1=1$; the top
  panels
are for $\Lambda_2=0.15$, while the bottom ones for
$\Lambda_2=1$. The dashed lines in the
left panels correspond to analytical predictions
(\ref{disp1})-(\ref{disp3}). Recall that instabilities emerge
when $\mathrm{Im}(\omega) \neq 0$.}\label{Fig1}
\end{figure}

For $\Lambda_2\leq0.51$, stable solutions appear only within a small
parametric interval near the AC limit. In this case, past
a ($\Lambda_2$ dependent) critical point an instability arises due to
 the collision of a localized eigenmode with the top of the dark
 component linear mode band, which leads to a Hopf bifurcation.
Additionally, the amplitude of the bright spot decreases as  $C$ is
increased (for fixed $\Lambda_2$) vanishing above a critical value of
$C$. For
small values of $\Lambda_2$, the solution vanishing takes place before
the
destabilization, implying that the DB soliton is stable in the whole
range of $C$ in which it exists with a nontrivial bright component.
Fig.~\ref{Fig1}~(top) shows for a prototypical value of $\Lambda_2$
within this parameter range, the dependence of the eigenmode spectrum
with respect to $C$.
The figure also features the comparison of the relevant
eigenfrequencies with
the analytical predictions (\ref{disp1})-(\ref{disp3}) showing an
excellent agreement for small $C$. Recall that per the above
analytical
predictions there is an internal mode (pertaining to the ``dark part''
of the DB object) which collides at finite $C$ with the expanding
band of phonon spectrum growing away from zero. This fact leads to the
observed
instability of the DB solitary wave past the relevant critical point
in $C$. When $\Lambda_2$ is increased, as it is evident from both
Fig.~\ref{Fig4}
and the bottom panels of Fig.~\ref{Fig1}, the critical $C$ for the
instability emergence tends to $0$.

\begin{figure}
\begin{center}
\includegraphics[width=0.6\textwidth]{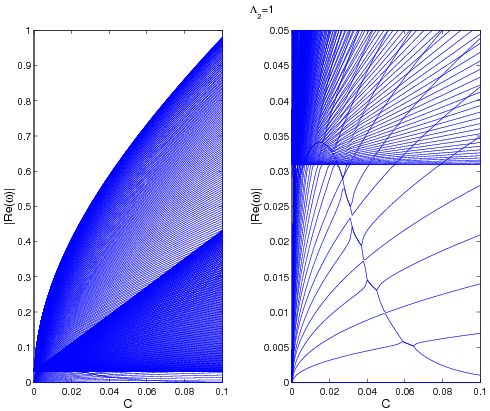}
\end{center}
\caption{Real part of eigenfrequencies versus coupling $C$ for dark-bright one-soliton solutions with $\Lambda_2=\Lambda_1=1$. The right panel is a zoom of the left panel where the the existence of a Hopf bifurcation cascade is highlighted.}\label{Fig2}
\end{figure}

\begin{figure}
\begin{center}
\includegraphics[width=0.6\textwidth]{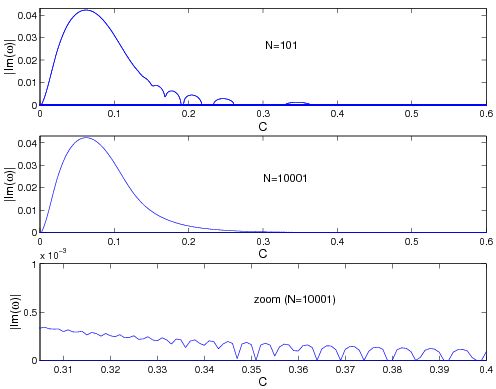}
\end{center}
\caption{The imaginary part of the eigenfrequencies is plotted
 versus the coupling $C$ for $\Lambda_2=0.9$. The dependence on the
 system size is illustrated: $N=101$ is chosen in the top panel whereas
$N=10001$ is taken in the other ones. The bottom panel highlights a
zoom-in of the middle panel
into a parametric range where bubbles appear.} \label{Fig3}
\end{figure}

\begin{figure}
\begin{center}
\includegraphics[width=0.6\textwidth]{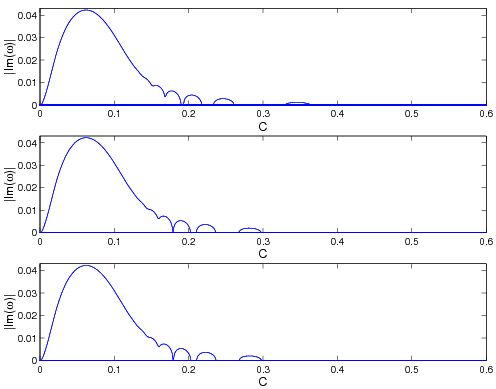}
\end{center}
\caption{The imaginary part of the eigenfrequencies is plotted
versus the coupling $C$ for $\Lambda_2=0.9$ and $N=101$. The (in)dependence
of  the  stability bubbles on the boundary conditions is highlighted.
The top panel corresponds to free end boundary conditions, the middle
panel to anti-periodic boundary conditions and fixed ends are shown in the
bottom panel. Notice that the imaginary parts of the eigenvalues are almost
identical for fixed ends and anti-periodic boundary conditions (the
differences are $\lesssim10^{-4}$.)} \label{Fig3b}
\end{figure}

For $\Lambda_2\in(0.51,1]$ solutions exist for every value of
$C$. Now, once the localized mode collides with the linear mode band,
new parametric
regions of stabilization appears as ``bubbles'' caused by a Hopf
bifurcation
cascade, as shown in Fig.~\ref{Fig2}. The size of the instability
bubbles
decreases and the number of them increases as the number of
particles $N$
increases. This is the so called boundary-induced stabilization and is
a
finite size effect discussed e.g. in ~\cite{joha}. Figure~\ref{Fig3}
illustrates, as an example, two stability diagrams obtained with
$N=101$ and
$N=10001$. The instability bubbles' size is almost independent
of the boundary conditions used, whereas the number of bubbles is
independent of them, as shown in Fig.~\ref{Fig3b}.
Fig.~\ref{Fig1}~(bottom) shows the eigenmodes spectrum for
an
example within this range.  This behaviour resembles to that of a dark soliton \cite{joha}.
Contrary to the above mentioned case, in this parametric regime,
there is no vanishing of the
bright spot and genuine DB solutions exist for every value of $C$.
Finally, for $\Lambda_2>1$, the DB soliton is unstable for every value of $C$.

\subsubsection{Dark-bright two-soliton solutions}

\begin{figure}
\begin{center}
\includegraphics[width=0.7\textwidth]{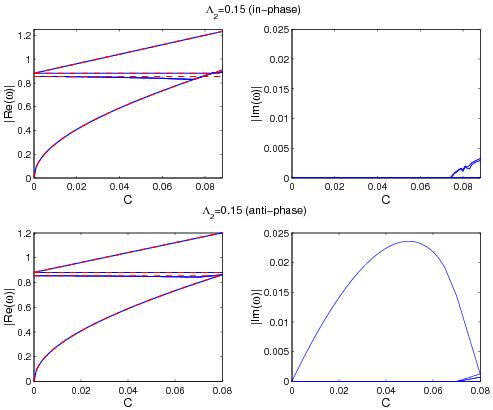}
\end{center}
\caption{Real part of the relevant eigenfrequencies (left panels) and imaginary part of the full spectrum (right panels) versus $C$ of the DB two-soliton solutions. Dashed lines in left panels correspond to analytical predictions (\ref{disp1})-(\ref{disp3}).}\label{Fig6}
\end{figure}

\begin{figure}
\begin{center}
\includegraphics[width=0.6\textwidth]{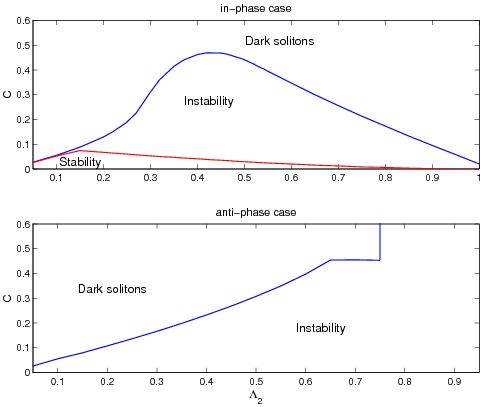}
\end{center}
\caption{Existence and  stability properties of the dark-bright
  two-soliton solutions as a function of coupling $C$ and frequency
  $\Lambda_2$, with the fixed value $\Lambda_1=1$. The top panel shows
the in-phase case, while the bottom one illustrates the anti-phase
case.}
\label{Fig5}
\end{figure}

Here we consider both the {\it in-phase} and {\it anti-phase} cases.

In the in-phase case, we observe that the DB solitons
behave structurally in a similar way to the case of the single
dark-bright
soliton
solutions with  $\Lambda_2\leq 0.51$ (left region of Fig.~\ref{Fig4}),
that is, for small $C$ the solutions are stable up to a critical value
where they destabilize, with their bright component disappearing above
a second critical value of $C$; additionally, stability bubbles do not
exist. Similarly, there do not exist regions where the DB pair
solution
persists
for arbitrary $C$. The latter is natural to expect since the in-phase
structure induces repulsion among the bright
components\footnote{This is an
important structural characteristic of the self-defocusing nature of our
nonlinearity. Namely, contrary to what is standardly known for
the case of self-focusing/attractive interactions \cite{yuri},
bright solitons here {\it repel} when they are {\it in phase},
but {\it attract} when they are {\it out of phase}; cf. also with the
continuum case discussion of Ref. \cite{berloff}.}; this fact, combined
with the repulsion among the dark components, does not allow the
possibility
of sustaining stationary such solutions near the continuum limit.
Moreover, we observe that the maximum value of $C$ does not depends
monotonically with $\Lambda_2$, having a maximum around
$\Lambda_2=0.42$. For $\Lambda_2$ below this critical value,
and upon increase of $C$, the
bright spot amplitude vanishes similar to the DB 1-S case but now the
solution that exists above the critical value of $C$
consists of 2 dark soliton waveforms; however, when $\Lambda_2$ is
above the critical value, the continuation fails past a
certain critical $C$. This is due to the existence of an inverse
Hopf bifurcation, which  may be caused by the interaction between
the two peaks of the bright component, similarly to
what is observed in \cite{varsol}.

On the contrary, the out-of-phase case solutions are {\it always}
exponentially unstable. Additionally, we observe a monotonically
increasing
behaviour of the maximum value of $C$ with $\Lambda_2$ up to
$\Lambda_2\approx0.65$. In this parametric range, the bright spot
vanishes
above the critical coupling. On the other hand, for
$0.65\lesssim\Lambda_2\lesssim0.75$, the curve is almost horizontal
and the
solutions experience an inverse Hopf bifurcation. Finally, DB 2-S exist for any value of $C$ whenever $\Lambda_2\geq0.75$.
These findings are summarized in Figs. \ref{Fig6} and \ref{Fig5}.
Let us mention in passing that somewhat surprisingly, the stability
for
small $C$ is the opposite than for the case of bright solitons which
are separated by an even number of sites (i.e. in the latter case, the
bright 2-solitons in-phase are unstable for any coupling whereas the
bright 2-solitons in anti-phase are stable for small coupling) \cite{varsol};
for a corresponding discussion in the multi-soliton interactions
in the 2d case, see e.g. \cite{pgk01}.

\subsubsection{Dark-bright and dark two-soliton solutions}

\begin{figure}
\begin{center}
\includegraphics[width=0.6\textwidth]{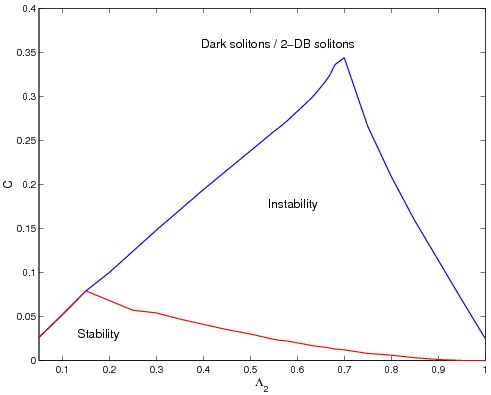}
\end{center}
\caption{Existence and  stability properties of the dark-bright and
dark two-soliton solutions as a function of coupling $C$ and
frequency $\Lambda_2$, with the fixed value $\Lambda_1=1$.}
\label{Fig7}
\end{figure}

The dark-bright and dark two-soliton solutions behave structurally in a similar way to
the DB 1-S or DB 2-S in-phase solutions. There are however some
important differences. The maximum value of $C$ (as a function
of $\Lambda_2$) for which such
solutions exist has a maximum for $\Lambda_2=0.70$. Below this value,
the
amplitude of $v_1<0$ (i.e. the amplitude of the absent bright spot),
increases its absolute value, and, above the maximum $C$ the DB+D
soliton transform into a DB 2-soliton solution in anti-phase. For
$\Lambda_2>0.70$, the soliton experiences an inverse Hopf bifurcation
at the
critical $C$. Figure~\ref{Fig7} illustrates the existence and
stability
properties of the dark-bright and dark two-soliton solutions as a
function of
the  coupling $C$ and the
frequency $\Lambda_2$.

\subsubsection{Summary of existence and stability results}

The previous findings for dark-bright two-soliton solutions  and for
the dark-bright and dark two-soliton pair solutions were obtained for a
separation of two sites between holes or peaks. However, the results
are similar when the separation increases up to 3 sites. The relevant
effects on the lattice decay exponentially with the separation, so
that for large values of this separation, the DB structures behave as
isolated ones and the stability properties are similar to the
dark-bright one-soliton solutions.

Arguably, the most interesting of the above results concerns the
stability
of the DB two-soliton state and especially the independence of this
result on
 the distance between the DB waves; instead, the result seems to
 solely (and {\it critically}) depend on the relative phase of the
 bright components. It is relevant to contra-distinct this to the case
 of simply two-dark solitons of the form defined above
 which would be stable (at least
close to the AC limit) independently on
 the distance $d$, and also to the case of two bright {\it gap}
 solitons,
considered before in \cite{pgk1} (see also \cite{pgk}) whose stability
would depend on {\it both} the distance $d$ and the relative phase. In
fact, in the
latter case of two bright gap solitons for defocusing nonlinearities,
using the so-called staggering transformation to convert the
defocusing dynamics into focusing one, it is straightforward to
establish~\cite{pgk1} that in-phase excited sites, separated by an
even
number of sites are unstable, while by an odd number of sites are
stable (for small $C$), while the situation is reversed for anti-phase
excitations (stable for an even site distance and unstable for an odd
one). On the other hand, here
we observe that the ``dressing'' of the
interaction of the bright components via their dark partners leads to
stability for in-phase excitations and instability for out-of-phase
ones. At a qualitative level, one can present the following
straightforward
argument towards understanding this effect: it is well-known (see also
the discussion below regarding DB soliton collisions) that DB solitons
with in-phase
bright components {\it repel} each other while DB states with
out-of-phase
bright excitations attract each other; see e.g. the fundamental
experimental work of \cite{seg2}, and the very recent theoretical
discussion
 of \cite{victor}. As a result the
state with out-of-phase bright excitation will be structurally less robust e.g. if the DB pair is moved closer, the attraction becomes
stronger and the tendency is to deviate further from the equilibrium. If the pair is moved further away, the attraction
becomes weaker, and again the further deviation from equilibrium is favored. The contrary is true for the in-phase excitation. The
repulsion of the bright components, if they are brought closer or
drift away from the equilibrium, favors in both cases their return to
equilibrium indicating its structural robustness.

\section{Collisions}\label{sec:collisions}

Having examined the stability of the DB soliton configurations, we
now turn to a more systematic examination of collision events
involving two DB waves or a dark-bright and a dark one. All the
numerical computations below have been performed perturbing only the
bright parts of the corresponding solutions. In the case of
dark-bright and dark two-soliton pairs, the perturbation only affects
the single bright spot, and for the other cases the perturbations
affect symmetrically each bright wavepacket so that they can move in opposite
directions. For most of the collisions we have taken $N=402$,
$\Lambda_1=1$, $\Lambda_2= 0.7$ and for each value of $C$, the
values of  $\alpha$ (the parameter controlling the ``kick''
$\propto e^{i \alpha (n-n_0)}$, where $n_0$ is the center of the
soliton), have been taken in the interval
$[0.01,0.30]$. The velocity of the dark-bright soliton increases when
$\alpha$ increases. The solitons are initially separated by
a distance of 200 lattice sites. Notice that only collisions at
intermediate and high values of $C$ have been considered, as
for low values of $C$ the solitons are not mobile due to
discreteness effects which impede their traveling. We have observed that, for $\Lambda_2=0.7$, which is the case analyzed in most of the cases below, $C$ must be higher than 1.2 if $\alpha=0.01$ and 0.9 if $\alpha=0.3$. The critical coupling needed for mobility decreases with $\Lambda_2$ and $\alpha$.

\subsection{Collisions between a dark-bright and a dark soliton}

\begin{enumerate}

\item{\em Collisions at intermediate coupling}

First of all, we consider collisions for $C=1.23$. The possible outcomes  of such events are illustrated in
Figs.~\ref{Fig10}-\ref{Fig11}, obtained with increasing values of $\alpha$. This value of $C$ has been chosen so
that it is ``intermediate'' between the small coupling regime (of low
soliton
mobility) and the high coupling regime where the collisions are essentially tantamount to the ones in the continuum model (and
discreteness is irrelevant). In this intermediate range, the discreteness plays a crucial role in determining the collisional outcome and provides a rich variety of potential scenarios. The collision events are illustrated by means of space-time diagrams representing the contour plots of the densities of the two fields, \{$|u_n|^2$\} and \{$|v_n|^2$\}, respectively.

For $\alpha$ small enough the static dark remains at rest after the
collision, and the dark-bright is reflected from it, as shown in
Fig.~\ref{Fig10} for $\alpha=0.08$. For larger values of $\alpha$
such as $0.12$ of the middle panel of Fig.~\ref{Fig10}, the
dark-bright may rebound but also concurrently leads to the motion
alongside it of the stationary dark soliton, formulating an
interesting co-propagating dark-bright and dark soliton pair. This
apparent effect of discreteness is not only absent for higher
couplings $C$ (as we will see below), but also is especially
unexpected given the apparently repulsive nature of the interaction between
the dark-bright and the dark soliton for large $C$ (i.e., in the continuum
limit), discussed below. This scenario appears for a small interval
of $\alpha$ values. The explanation that we offer for this effect is
the following: for the present value of $C$ and the relevant lattice
size $N=402$, the dark soliton is, in fact, {\it dynamically
unstable}. While this instability does not manifest itself until the
collision event, the latter appears to {\it trigger} the instability
by providing a significant perturbation to the stationary dark
soliton. As a result, after the collision the dark soliton becomes
subject to the oscillatory instability, in turn giving rise (as observed
previously) to its motion; see also~\cite{oksana}.

For increasing values of $\alpha$
the static dark always gets into motion but some different scenarios
appear:
\begin{enumerate}
\item

 The dark-bright solitary wave rebounds launching out away the static dark, as
illustrated for $\alpha=0.13$ in Fig.~\ref{Fig10}.
 \item

The dark-bright wave may remain nearly stopped, as for
$\alpha=0.24$ in the left panel of Fig.~\ref{Fig11}.

\item The dark-bright may continue traveling along the incident direction,
as the dark (initially stationary) soliton does; an example of this
 for $\alpha=0.245$ can be found in the middle panel of Fig.~\ref{Fig11}.

\item
Lastly, we observe another unusual dynamical  scenario unfold in the
right panel of Fig.~\ref{Fig11}: here, apparently the dynamical
instability of the stationary dark soliton manifests itself prior to
the arrival of the DB soliton (and the ensuing collision event).
As a result, after the collision the
dark-bright continues traveling without essentially changing its velocity
while the dynamically unstable dark soliton continues traveling
in the opposite direction.

\end{enumerate}
\begin{figure}[ht]
\centerline{\includegraphics[width=0.7\textwidth]{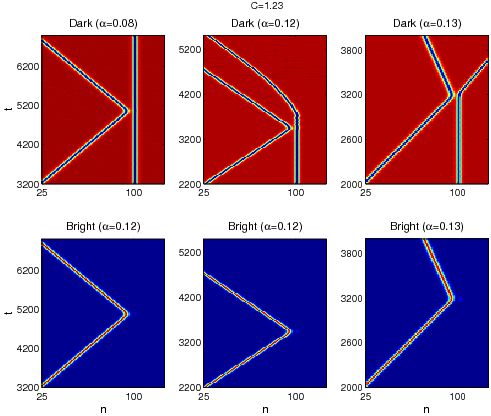}}
\caption{Three dark-bright+dark 2-S collisions corresponding to
increasing values of $\alpha$; $\alpha=0.08$ on the left,
$\alpha=0.12$
in the middle and $\alpha=0.13$ on the right.
Relevant parameters are chosen as:
$C=1.23$; $\Lambda_1=1$; $\Lambda_2=0.7$.}
\label{Fig10}
\end{figure}

\begin{figure}[ht]
\centerline{\includegraphics[width=0.7\textwidth]{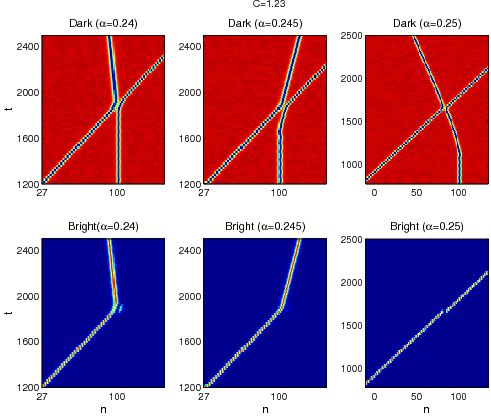}}
\caption{Same as Fig.\ref{Fig10}, but for larger values of the
initial kick $\alpha$ ($0.24$, $0.245$ and $0.25$, respectively on the
left, middle and right), corresponding to larger speeds.}
\label{Fig11}
\end{figure}

\item{\em Collisions at higher coupling }

Some prototypical examples of collisions at higher coupling are also
given for $C=3$ in Fig. \ref{Fig12}. These are essentially for a
regime where discreteness is irrelevant and the dynamics chiefly
represents the near-elastic collisions of the continuum limit
(which, however, can still be fairly complex, exhibiting beating and
breathing phenomena as illustrated in \cite{aravi}). Nevertheless, we
show them here to showcase the repulsive nature of the interaction
between the dark-bright and the dark soliton. It is clear that the
low velocity collision of the left panel of the figure signals a
repulsive occurrence whereby the centers of the two solitons never
``touch''. There is a critical kinetic energy which enables the DB
soliton to balance the  repulsion induced from the stationary dark
soliton. Close to this critical point, there is a prolonged merger
of the two waves, prior to their separation as evident in the
middle panel of the figure. Finally, the right panel shows a
supercritical case, where the DB soliton possesses enough kinetic
energy to {\it overcome} the repulsive barrier posed by the dark
stationary state. In that case, the dark remains stationary, while
the dark-bright passes through it. It should be noted that in these
events, the large value of $C$ mitigates the dynamical instability
(which now has a very small growth rate as the continuum limit is
approached), hence some of the more elaborate occurrences of Figs.
~\ref{Fig10}-\ref{Fig11} are absent here.

\begin{figure}[ht]
\centerline{\includegraphics[width=0.7\textwidth]{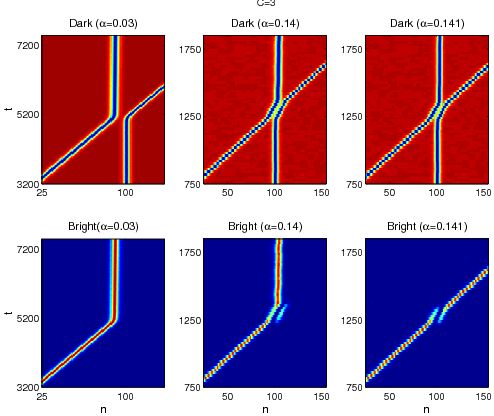}}
\caption{Dark-bright and dark solitary wave collisions corresponding
to increasing values of $\alpha$; $\alpha=0.03$,
$0.14$ and $0.141$, respectively for the left, middle and right
panels. Other parameters are chosen as $C=3$; $\Lambda_1=1$;
$\Lambda_2=0.7$.}
\label{Fig12}
\end{figure}

\end{enumerate}

\subsection{Two dark bright solitons collisions}

In order to observe the collision scenario between two DB solitons, we
fix $C=1.55$. This value of $C$ has been chosen so that it is
``intermediate'', as in the
above examined collisions between a dark-bright and a dark soliton.

First of all, we consider slow soliton collisions. The parameter $\alpha$ should be small enough so that the kinetic
energy does not overwhelm the details of the collisions. For these
low velocities dark-bright solitons with in-phase bright components
{\it repel} each other while dark-bright solitons with out-of-phase
bright excitations attract each other. These results agree with the
experimental work of \cite{seg2} and the theoretical work of
\cite{berloff} which discusses the nature of the interaction.

\begin{figure}[ht]
\centerline{\includegraphics[width=0.7\textwidth]{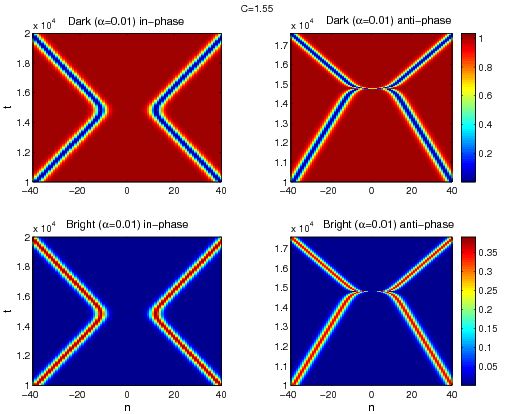}}
\caption{An example of  dark-bright two-soliton collisions. (Left):
in-phase case. (Right): anti-phase case. Parameters: $C=1.55$;
$\Lambda_1=1$; $\Lambda_2=0.7$ and $\alpha=0.01$.}
 \label{Fig8}
\end{figure}

As an example, taking $\alpha=0.01$, Fig.~\ref{Fig8} illustrates a dark-bright two-soliton collision event for the in-phase case (left column) and for the anti-phase case (right column), with  $\Lambda_2=0.7$. We observe repulsion of the dark components when the bright components are in phase and attraction when they are out of phase.

\begin{figure}[ht]
\centerline{\includegraphics[width=0.7\textwidth]{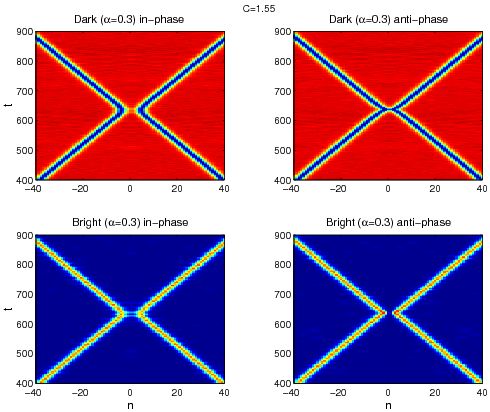}}
\caption{Same as Fig. \ref{Fig8}, but for $\alpha=0.3$ that
corresponds to a larger speed.}
 \label{Fig9}
\end{figure}

If the velocity is increased, the following modifications are
observed.
On the one hand, in the  in-phase bright component case, the repulsion
of both
the dark and bright components is mitigated by the increase of kinetic
energy (i.e., the turning points of the relative motion occur closer
to $n=0$); on the other hand, for the anti-phase bright component,
the kinetic energy and attractive interaction cooperate in bringing
the solitons together.
Fig.~\ref{Fig9} shows an example of the outcome of both cases
for $\alpha=0.3$.

\begin{figure}[ht]
\centerline{\includegraphics[width=0.7\textwidth]{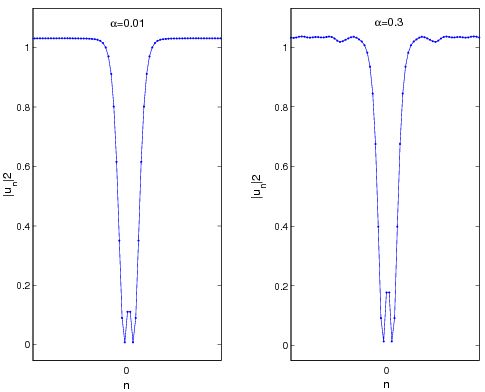}}
\caption{Density profile of the dark components of the colliding solitons of right panels of Fig.~\ref{Fig8} (left) and Fig.~\ref{Fig9} when the distance between them is the shortest.}
\label{Figcol}
\end{figure}

In order to discern if the dark components of the two DBs
with $\pi$ out of phase bright components
cross each other or not, we have traced in Fig.~\ref{Figcol} the dark
component density profile $|u_n|^2$ at the time where both colliding
solitons are at the minimum distance for the right panels of
Figs.~\ref{Fig8} and \ref{Fig9}. We can observe the existence of two
dips instead of a single one, which is a clear indication
that the dark components  of the solitons do not cross each other.
In fact, as discussed also in \cite{josab}, it is only when the
two dark soliton centers coincide that they pass through each
other (i.e., that they overcome the finite barrier of repulsion
between them).

\begin{figure}[ht]
\centerline{\includegraphics[width=0.7\textwidth]{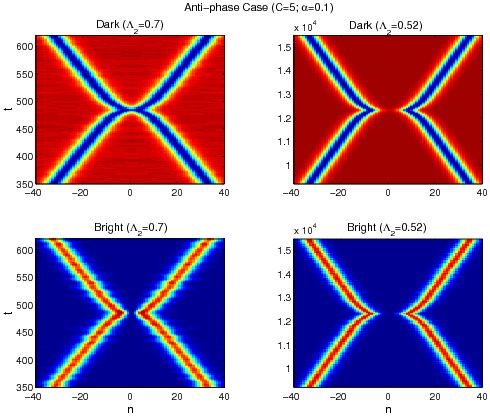}}
\caption{Comparison between 2 DB soliton collisions when $\Lambda_2$
is varied ($\Lambda_2=0.7$ on the left, while $\Lambda_2=0.52$ on the right; $C=5$ and $\alpha=0.1$ in both cases).
It is observed that below a critical value, the dark component
repulsion
can be deemed to be stronger than the anti-phase bright component
attraction
and hence creates a potential barrier that, in turn, induces
a finite separation ``turning point'' in the interaction between
the solitary waves.}
\label{Fig13}
\end{figure}

Ref.~\cite{berloff}
considers the interaction between two DBs with $\pi$ out of phase
bright components as the norm of the bright component is modified.
There, the
fading attraction effect between the bright components
(as their norm decreases) enables the
emergence of a repulsive character of the interaction due to the
dark components.
In order to check if this phenomenon also occurs in our system, we
have performed some simulations varying $\Lambda_2$ (increasing
$\Lambda_2$ is equivalent
to increase the norm of the bright component) showing that there is a
critical value of $\Lambda_2$ below (above) which the dark components
repel
(attract) each other. This phenomenon is illustrated in Fig.~\ref{Fig13}.

Finally, we have checked that, contrary to the dark-bright and dark
solitary wave
collisions, no fundamental qualitative differences are observed when a high
coupling in implemented. That is, collisions at intermediate and
strong couplings $C$ roughly exhibit the same principal phenomenology
(hence, we do not show any of the latter here).

\section{Conclusions and Future Challenges}

In the present work, we offered a numerical perspective on the
existence, stability, dynamics and interactions of single
and multiple dark-bright solitons not only between them,
but also with dark solitons in the setting of nonlinear
dynamical lattices. The presence of discreteness  enabled a number
of interesting variants to the previously studied continuous
setting. Some of these variants could be anticipated
on the basis of analogous studies of dark solitons~\cite{pgk}.
E.g., the dark-bright single solitons could be stable only up to
a critical coupling (although finite-size effects could provide
additional stabilization). The same thing would be true for
pairs of dark-bright and dark solitons. The collisions of
the latter pairs indicate the repulsive interaction between
the dark-bright and the dark states. On the other hand, a number
of interesting novel features were revealed. Relevant examples
include the generic instability of the dark-bright for a sufficiently
strong bright component, the generic instability of out-of-phase
two-DB soliton pairs, and the generic stability (again for
large enough bright component) of in-phase two-DB soliton pairs.
Discreteness was also found to play a significant role in affecting
collisions for ``intermediate'' coupling strengths (not too small,
so that the excitations would get pinned, nor too large so that they
would interact in a quasi-continuum way). There, its potential combination
with the dynamical instability e.g. of stationary dark solitons could
yield a variety of interesting collisional outcomes, especially
in the collisions of dark-bright solitary waves with dark ones.

There are numerous directions in which one can consider expanding
the considerations presented herein. One possible example is an
analytical attempt to calculate the linearization eigenvalues
of two-DB soliton pairs, to explicitly prove the stability conclusions
numerically inferred herein (and to obtain some mathematical
intuition on these findings). Another possibility from a numerical
viewpoint is to extend the considerations presented herein to
vortex-bright solitons (or to dark-bright rings)
in higher-dimensional settings, by
analogy to the continuum form of such symbiotic entities recently
considered in \cite{VB}. There one can potentially envision situations
where the symbiotic structure may be stable although neither of its
constituents is robust by itself.
Potential experimental realizations
of dark-bright solitons and soliton pairs in waveguide arrays
(such as lithium niobate ones) could be an excellent source
of testing of the relevant ideas in realistic settings.

\section*{Acknowledgments}

PGK gratefully acknowledges support from
NSF-DMS-0349023 (CAREER), NSF-DMS-0806762 and the Alexander von
Humboldt Foundation. AA, FRR and JC acknowledge financial support
from the MICINN project FIS2008-04848.


\begin{thebibliography}{99}
\bibitem{yuri} Yu.S. Kivshar and G.P. Agrawal,
{\it Optical solitons: from fibers to photonic crystals},
Academic Press (San Diego, 2003).

\bibitem{stringarius} P. G. Kevrekidis, D. J. Frantzeskakis, and R.
Carretero-Gonz{\'a}lez, {\it Emergent Nonlinear Phenomena in
Bose-Einstein Condensates}, Springer-Verlag (Berlin, 2008).

\bibitem{luther} Yu.S. Kivshar and B. Luther-Davies,
Phys. Rep. {\bf 298}, 82 (1998).

\bibitem{seg1} Z. Chen, M. Segev, T.H. Coskun, D.N. Christodoulides,
Yu.S. Kivshar and V.V. Afanasjev, Opt. Lett. {\bf 21},
1821 (1996).

\bibitem{seg2} E.A. Ostrovskaya, Yu.S. Kivshar, Z. Chen and M. Segev,
Opt. Lett. {\bf 24},
327 (1999).

\bibitem{ysk91} Y.S. Kivshar, D. Anderson, A. H{\"o}{\"o}k,
M. Lisak, A.A. Afanasjev and V.N. Serkin,
Phys. Scripta {\bf 44}, 195 (1991).

\bibitem{Myatt1997a} C.J. Myatt,
E. A. Burt, R. W. Ghrist, E. A. Cornell, and C. E. Wieman
Phys. Rev. Lett. {\bf 78}, 586 (1997).

\bibitem{Hall1998a} D. S. Hall,
M. R. Matthews, J. R. Ensher, C. E. Wieman, and E. A. Cornell
Phys. Rev. Lett. {\bf 81}, 1539 (1998).

\bibitem{Stamper-Kurn1998b} D.M. Stamper-Kurn,
M. R. Andrews, A. P. Chikkatur, S. Inouye,
H.-J. Miesner, J. Stenger, and W. Ketterle,
Phys. Rev. Lett. {\bf 80}, 2027 (1998).

\bibitem{buschanglin} Th. Busch and J.R. Anglin,
Phys. Rev. Lett. {\bf 87}, 010401 (2001).

\bibitem{DDB} H.E. Nistazakis, D.J. Frantzeskakis, P.G. Kevrekidis,
B.A. Malomed and R. Carretero-Gonz{\'a}lez,
Phys. Rev. A {\bf 77}, 033612 (2008).

\bibitem{rajendran} S. Rajendran, P. Muruganandam,
M. Lakshmanan, J. Phys. B {\bf 42}, 145307 (2009).

\bibitem{berloff}  C. Yin, N.G. Berloff, V.M. P\'erez-Garc\'{\i}a, V.A. Brazhnyi
and H. Michinel,
arXiv:1003.4617

\bibitem{VB} K.J.H. Law, P.G. Kevrekidis and L.S. Tuckerman,
Phys. Rev. Lett. {\bf 105}, 160405 (2010).

\bibitem{hamburg} C. Becker, S. Stellmer, P. Soltan-Panahi,
S. D{\"o}rscher, M. Baumert, E.-M. Richter, J. Kronj{\"a}ger, K. Bongs,
and K. Sengstock,
Nature Phys. {\bf 4}, 496 (2008).

\bibitem{enge1} C. Hamner, J.J. Chang, P. Engels, M. A. Hoefer,
arXiv:1005.2610.

\bibitem{enge2} S. Middelkamp, J.J. Chang, C. Hamner,
R. Carretero-Gonzalez, P.G. Kevrekidis, V. Achilleos,
D.J. Frantzeskakis, P. Schmelcher, P. Engels,
arXiv:1005.3789.


\bibitem{efrem} N. K. Efremidis,
S. Sears, D. N. Christodoulides, J. W. Fleischer, and M. Segev
Phys. Rev. E {\bf 66}, 46602 (2002).

\bibitem{moti1} J. W. Fleischer, M. Segev, N. K. Efremidis, and
D. N. Christodoulides, Nature {\bf 422}, 147 (2003).

\bibitem{moti2} J. W. Fleischer, T. Carmon,
M. Segev, N. K. Efremidis, and D. N. Christodoulides,
Phys. Rev. Lett. {\bf 90}, 23902 (2003).

\bibitem{lederer} F. Lederer, G.I. Stegeman, D.N. Christodoulides,
G. Assanto, M. Segev and Y. Silberberg, Phys. Rep. {\bf 463}, 1 (2008).

\bibitem{gorbach} S. Flach and A.V. Gorbach,
Phys. Rep. {\bf 467}, 1 (2008).


\bibitem{7} H.S. Eisenberg, Y. Silberberg,
R. Morandotti, A. R. Boyd and J. S. Aitchison, Phys. Rev. Lett. {\bf 81}, 3383 (1998).

\bibitem{review_opt} D. N.\ Christodoulides,
F.\ Lederer, and Y.\ Silberberg, Nature \textbf{424}, 817 (2003).

\bibitem{review_opt1} A. A.\ Sukhorukov,
Yu. S.\ Kivshar, H. S.\ Eisenberg, and Y.\ Silberberg, IEEE J. Quant. Elect. \textbf{39}, 31 (2003).

\bibitem{ibloch} I. Bloch, Nature Phys. {\bf 1}, 23 (2005).


\bibitem{konotop} V. A. Brazhnyi and V. V. Konotop, Mod. Phys. Lett. B
\textbf{18}, 627 (2004); P. G. Kevrekidis and D. J. Frantzeskakis, Mod.
Phys. Lett. B \textbf{18}, 173 (2004).

\bibitem{markus2} O. Morsch and M. Oberthaler, Rev. Mod. Phys. {\bf 78}, 179 (2006).

\bibitem{silb} R. Morandotti, H. S. Eisenberg, Y. Silberberg,
M. Sorel, and J. S. Aitchison, Phys. Rev. Lett. {\bf 86}, 3296 (2001).

\bibitem{kip} E. Smirnov, C. E. R{\"u}tter, M. Stepi{\'c}, D. Kip, and V. Shandarov,
Phys. Rev. E {\bf 74}, 065601 (2006).

\bibitem{christo} J. Meier, J. Hudock, D.N. Christodoulides, G. Stegeman,
Y. Silberberg, R. Morandotti, and J. S. Aitchison,
Phys. Rev. Lett. {\bf 91}, 143907 (2003).

\bibitem{mackay} R.S. MacKay and S. Aubry,
Nonlinearity {\bf 7}, 1623 (1994).

\bibitem{pgk} P.G. Kevrekidis, {\it The discrete nonlinear
Schr{\"o}dinger equation: mathematical analysis, numerical computations
and physical perspectives}, Springer-Verlag (Heidelberg, 2009).

\bibitem{Interlaced} J. Cuevas, Q.E. Hoq, H. Susanto, and P. G. Kevrekidis.
Physica D {\bf 238}, 2216 (2009); J. Cuevas, G. James, P.G. Kevrekidis and K.J.H. Law.
Physica D {\bf 238}, 1422 (2009).

\bibitem{joha} M. Johansson and Yu.S. Kivshar,
Phys. Rev. Lett. {\bf 82}, 85 (1999).

\bibitem{varsol} J. Cuevas, B.A. Malomed, D.J. Frantzeskakis and P.G. Kevrekidis. Physica D {\bf 238}, 67 (2009).

\bibitem{pgk01} P.G. Kevrekidis, B.A. Malomed and A.R. Bishop,
J. Phys. A: Math. Gen. {\bf 34}, 9615 (2001).

\bibitem{pgk1} P. G. Kevrekidis, H. Susanto, and Z. Chen,
Phys. Rev. E {\bf 74}, 066606 (2006).

\bibitem{victor} V.A. Brazhnyi and V.M. P\'erez-Garc\'{\i}a,
arXiv:1004.3672.

\bibitem{oksana} Yu.S. Kivshar, W. Kr\'olikowski and O.A. Chubykalo,
Phys. Rev. E {\bf 50}, 5020 (1994).

\bibitem{aravi} R. Radhakrishnan and K. Aravinthan,
J. Phys. A: Math. Theor. {\bf 40}, 13023 (2007).


\bibitem{josab} R.N. Thurston and A.M. Weiner,
J. Opt. Soc. Am. B {\bf 8}, 471 (1991).

\end{thebibliography}
\end{document}